\documentclass{jpp}

\usepackage{amsmath}

\newcommand{\eq}[1]{(\ref{#1})}
\newcommand{\Eq}[1]{Eq.~(\ref{#1})}
\newcommand{\Eqs}[1]{Eqs.~(\ref{#1})}
\newcommand{\Sec}[1]{Sec.~\ref{#1}}
\newcommand{\mc}[1]{\mathcal{#1}}
\newcommand{\mcc}[1]{\mathfrak{#1}}
\renewcommand{\vec}[1]{{\boldsymbol{#1}}}
\newcommand{\pd}{\partial}
\newcommand{\ui}{{\rm i}}
\newcommand{\ue}{{\rm e}}

\shorttitle{Photon polarizability and its effect on the dispersion of plasma waves}
\shortauthor{I. Y. Dodin and D. E. Ruiz}

\title{Photon polarizability and its effect on the dispersion of plasma waves}

\author{I. Y. Dodin$^{\boldsymbol{1,2}}$
  \corresp{\email{idodin@princeton.edu}}
 \and D. E. Ruiz$^{\boldsymbol{1}}$}

\affiliation{
\aff{1}Department of Astrophysical Sciences, Princeton University, Princeton, New Jersey 08544, USA
\aff{2}Princeton Plasma Physics Laboratory, Princeton, New Jersey 08543, USA
}

\begin{document}

\maketitle

\begin{abstract}
High-frequency photons traveling in plasma exhibit a linear polarizability that can influence the dispersion of linear plasma waves. We present a detailed calculation of this effect for Langmuir waves as a characteristic example. Two alternative formulations are given. In the first formulation, we calculate the modified dispersion of Langmuir waves by solving the governing equations for the electron fluid, where the photon contribution enters as a ponderomotive force. In the second formulation, we provide a derivation based on the photon polarizability. Then, the calculation of ponderomotive forces is not needed, and the result is more general.
\end{abstract}


\section{Introduction}

As we showed recently \citep{tex:myqponder}, high-frequency photons traveling in plasma exhibit a well defined linear polarizability. Hence, they contribute to the linear dielectric tensor just like any other plasma particles, such as electrons and ions. This implies that high-frequency photons can influence the dispersion of linear plasma waves. Here, we present a detailed calculation of this effect for Langmuir waves~(LW).

Specifically, we develop a theory linear with respect to the LW amplitude $\varepsilon$. The photon density is assumed $O(\varepsilon^0)$, and perturbations to the photon density are assumed $O(\varepsilon^1)$. Hence, the LW can be understood as the linear modulational dynamics of the electromagnetic (EM) radiation. Two alternative formulations of this dynamics are given. In the first formulation (\Sec{sec:pde}), we calculate the LW dispersion by solving the governing equations for the electron fluid, where the photon contribution enters as a ponderomotive force. A related calculation was also reported previously \citep{ref:bingham97, book:mendonca}, but it has omissions that warrant a reconsideration. In the second formulation (\Sec{sec:pol}), we invoke the photon-polarizability concept. Then, the theory becomes \textit{linear}, ponderomotive forces do not need to be considered, and,  consequently, more general results are obtained.

Although we focus on LW, the calculation presented here is only a characteristic example. The concept of the photon (plasmon, phonon, etc.) polarizability can be useful also in more general settings. For example, effects related to those considered here also apply to waves in solid-state media \citep{ref:dylov08}. We leave the consideration of such specific examples (other than LW) to future publications. The present paper aims only to illustrate the basic idea.

\section{PDE-based approach}
\label{sec:pde}

In this section, we present a formulation that is based on the partial-differential equations (PDE) governing the plasma motion. We assume that, to the zeroth order in $\varepsilon$, the plasma (including the photon content) is stationary and homogeneous. We also assume that the ions are motionless, that the electrons can be modeled as a fluid, and that the $O(\varepsilon^0)$ velocity of the electron fluid is zero. We also assume that the electrons are collisionless and nonmagnetized. Hence, the wave dynamics is described as follows.

Consider the electron continuity equation
\begin{gather}\label{eq:ce}
\pd_t n + \nabla \cdot (n \vec{v}) = 0,
\end{gather}
where $n$ is the electron density, and $\vec{v}$ is the electron flow velocity. After the linearization, \Eq{eq:ce} becomes
\begin{gather}\label{eq:lce}
\pd_t \tilde{n} + n_0\nabla \cdot \tilde{\vec{v}} = 0.
\end{gather}
[We use subscript 0 to denote $O(\varepsilon^0)$ quantities and tilde to denote $O(\varepsilon^1)$ quantities.] The velocity $\tilde{\vec{v}}$ is found from the electron momentum equation
\begin{gather}\label{eq:me}
m n (\pd_t + \vec{v} \cdot \nabla)\vec{v} = n e \vec{E} - \nabla P + \vec{\eta},
\end{gather}
where $m$ and $e<0$ are the electron mass and charge, $P$ is the pressure, and $\vec{\eta}$ is the ponderomotive force density (averaged over the EM-field oscillations yet not over the LW oscillations). After the linearization, \Eq{eq:me} becomes
\begin{gather}
\pd_t \tilde{\vec{v}} = \frac{e}{m}\, \tilde{\vec{E}} - \frac{\nabla \tilde{P}}{m n_0} + \frac{\tilde{\vec{\eta}}}{m n_0}.
\end{gather}
The electron gas is considered adiabatic, so $\tilde{P} = 3m v_T^2\tilde{n}$ [see, e.g., \citep{my:dense}], where the constant $v_T$ is the unperturbed thermal speed of electrons. Hence, we get
\begin{gather}
\pd_t\tilde{\vec{v}} = \frac{e}{m}\, \tilde{\vec{E}} 
- 3v_T^2\, \frac{\nabla \tilde{n}}{n_0} + \frac{\tilde{\vec{\eta}}}{m n_0}.
\end{gather}
Substituting this into \Eq{eq:lce} gives $\pd_t^2\tilde{n} = -n_0\nabla \cdot \pd_t\tilde{\vec{v}}$, or
\begin{gather}
\pd_t^2\tilde{n}
 = -\frac{e n_0}{m}\,\nabla \cdot  \tilde{\vec{E}}
+ 3v_T^2 \nabla ^2\tilde{n} - \frac{\nabla \cdot \tilde{\vec{\eta}}}{m}.
\end{gather}
Using Gauss's law, $\nabla \cdot \tilde{\vec{E}} = 4\upi e \tilde{n}$, we then obtain
\begin{gather}
\pd_t^2\tilde{n}+ \omega_{p,0}^2 \tilde{n}-3v_T^2\nabla^2\tilde{n}
=-\frac{\nabla \cdot \tilde{\vec{\eta}}}{m},
\end{gather}
where $\smash{\omega_{p}^2\doteq (4\upi n e^2/m)^{1/2}}$. (We use the symbol $\doteq$ to denote definitions.) Assume $\tilde{n} = \text{Re}\,(n_c \smash{\ue^{-\ui\Omega t + \ui \vec{K}\cdot \vec{x}}})$ and similarly for other ``tilded'' quantities. Then,
\begin{gather}\label{eq:epwe}
[\Omega^2-\Omega_0^2(\vec{K})]\tilde{n}_c
= \ui \vec{K}\cdot \tilde{\vec{\eta}}_c/m,
\end{gather}
where $\Omega_0(\vec{K}) = (\omega_{p,0}^2+3K^2 v_T^2)^{1/2}$ is the LW frequency absent photons. Note that ${K v_T \ll \Omega}$ is implied because otherwise electrons cannot be considered adiabatic but rather must be described kinetically \citep{book:stix}; hence,
\begin{gather}\label{eq:Om0}
\Omega_0(\vec{K}) \approx \omega_{p, 0}(1 + 3K^2v_T^2/2\omega_{p,0}^2).
\end{gather}

In the case of broad-band EM radiation, the average $\vec{\eta}$ equals the sum of the ponderomotive forces produced by its individual quasimonochromatic constituents, i.e., traveling geometrical-optics (GO) waves with well defined wave vectors $\vec{k}$ and the corresponding frequencies
\begin{gather}\label{eq:om}
\omega = (\omega_p^2 + c^2 k^2)^{1/2}.
\end{gather}
Each such wave produces a per-electron average force $-\nabla \Phi_{\vec{k}}$, where $\Phi_{\vec{k}} = e^2 |\vec{E}_c|^2/(4m\omega^2)$ is the ponderomotive potential, and $\vec{E}_c$ is the complex amplitude of the EM-wave electric field $\vec{E}$ \citep{ref:gaponov58}. [More specifically, we adopt $\vec{E} = \text{Re}\,(\vec{E}_c \smash{\ue^{\ui\theta}})$, where $\theta$ is the wave rapid phase. Accordingly, $\omega \doteq - \pd_t \theta$ and $\vec{k} \doteq \nabla \theta$.] Hence, the average force density is $\vec{\eta}_{\vec{k}} = - n \nabla \Phi_{\vec{k}}$. Let us also express this in terms of the wave action density $\mc{I} = \mc{E}/\omega$, where \citep{my:amc} 
\begin{gather}
\mc{E} = \frac{1}{16\upi\omega}\,\vec{E}_c^* \cdot
\pd_{\omega}[\omega^2\vec{\epsilon}(\omega, \vec{k})] \cdot \vec{E}_c
\end{gather}
is the wave energy density, and $\vec{\epsilon}$ is the dielectric tensor. Using $\vec{\epsilon}(\omega, \vec{k}) = 1 - \omega_p^2/\omega^2$, we get $\mc{E} = |\vec{E}_c|^2/(8\upi)$, so
\begin{gather}
\vec{\eta}_{\vec{k}} = - \frac{\omega_p^2}{2}\,\nabla\left(
\frac{\mc{I}}{\omega}
\right).
\end{gather}
The force density $\vec{\eta}$ produced by broad-band radiation can be written as $\vec{\eta} = \int \vec{\eta}_{\vec{k}}\,d^3k$, where $\mc{I}$ is replaced with the \textit{phase-space} action density $F$. [One can also understand $F/\hbar$ as the phase-space photon probability distribution \citep{my:wkin}.] This gives
\begin{gather}\label{eq:fE}
\vec{\eta}(t, \vec{x}) = - \frac{\omega_p^2}{2}\,\nabla \int
\frac{F(t, \vec{x}, \vec{k})}{\omega(t, \vec{x}, \vec{k})}\,d^3k.
\end{gather}

Equation \eq{eq:fE} is in agreement with the formula reported previously \citep{ref:bingham97, book:mendonca}, at least up to a factor of two. However, in this work, we report a different linearization, which is as follows:
\begin{gather}
\tilde{\vec{\eta}}_c = - \frac{\ui}{2}\,\vec{K}\omega_{p,0}^2 \int
\left(
\frac{\tilde{F}_c}{\omega_0} - \frac{\tilde{\omega}_c F_0}{\omega_0^2}
\right)
d^3k.\label{eq:etac}
\end{gather}
Here, $\omega_0 \doteq (\omega_{p,0}^2 + c^2 k^2)^{1/2}$ is the unperturbed frequency of the EM wave, and
\begin{gather}\label{eq:omc}
\tilde{\omega}_c = \frac{\omega_{p,0}^2}{2\omega_0}\,\frac{\tilde{n}_c}{n_0}
\end{gather}
is the perturbation on the EM-wave frequency due to the plasma density variations caused by the LW. The perturbation of the photon distribution $\smash{\tilde{F}_c}$ is obtained from the wave kinetic equation
\begin{gather}\label{eq:wke}
\pd_t F + \pd_{\vec{k}} \omega \cdot \nabla F - \pd_{\vec{x}}\omega \cdot \nabla_{\vec{k}} F = 0,
\end{gather}
where $\omega = \omega(t, \vec{x}, \vec{k})$, and $\pd_{\vec{k}} \omega$ is understood as the group velocity. Since the unperturbed plasma is considered homogeneous, linearizing \Eq{eq:wke} leads to
\begin{gather}\label{eq:Fc}
\tilde{F}_c = - \frac{\tilde{\omega}_c \vec{K}  \cdot \nabla_{\vec{k}} F_0}{\Omega - \vec{K} \cdot \vec{v}_*},
\end{gather}
where $\vec{v}_* \doteq c\vec{k}/\omega_0$ is the unperturbed group velocity. Inserting \Eqs{eq:omc} and \eq{eq:Fc} into \Eq{eq:etac}, we obtain
\begin{gather}\label{eq:ft}
\tilde{\vec{\eta}}_c
 = \frac{\ui}{4}\, \left( \frac{\tilde{n}_c}{n_0} \right) \vec{K}\omega_{p,0}^4 \, Q(\Omega, \vec{K}),
\end{gather}
where we introduced the following function:
\begin{gather}
Q(\Omega, \vec{K}) \doteq \int
\left[
\frac{\vec{K}  \cdot \nabla_{\vec{k}} F_0(\vec{k})}{(\Omega - \vec{K} \cdot \vec{v}_*)\omega_0^2(k)} + \frac{F_0(\vec{k})}{\omega_0^3(k)}
\right]
d^3k.\label{eq:Q}
\end{gather}
By substituting \Eq{eq:ft} into \Eq{eq:epwe}, we get
\begin{gather}\notag
\Omega^2-\Omega_0^2(\vec{K})
= -\frac{\omega_{p,0}^4K^2}{4m n_0}\,Q(\Omega, \vec{K}).
\end{gather}
Assuming that $\Omega$ is close to $\Omega_0\approx \omega_{p,0}$, we can also simplify the left side here as follows:
\begin{gather}
\Omega^2-\Omega_0^2(\vec{K})
\approx 2\omega_{p,0}[\Omega -\Omega_0(\vec{K})].
\end{gather}
Then, the dispersion relation becomes
\begin{gather}\label{eq:Omg}
\Omega \approx \Omega_0(\vec{K}) - \frac{\omega_{p,0}^3K^2}{8m n_0}\,Q(\Omega, \vec{K}).
\end{gather}
The second term on the right side is the modification of the LW dispersion relation caused by the photon gas. Equation \eq{eq:Omg} differs from the corresponding relation reported previously \citep{ref:bingham97, book:mendonca} in the following aspects: (a)~the order-one numerical coefficient in front of the integral is different, and, most importantly, the second term in the integrand in our expression for $Q$ [\Eq{eq:Q}] is new. Notably, such terms are missed also in the general method reported by \citet{book:tsytovich} for calculating the nonlinear plasma dispersion. The wave-scattering paradigm that underlies this method assumes that the interaction between each pair of waves can be modeled as instantaneous, so the effect of the adiabatic frequency shift $\omega - \omega_0$ cannot be captured.

In order to estimate the photon contribution, suppose for clarity that $\omega \sim k c$ and $k \sim K$. (Under the assumptions adopted in the present section, this regime is accessible only marginally, but a more general theory given in \Sec{sec:pol} leads to similar estimates.) Then, the ratio of the two terms in the right side of \Eq{eq:Omg} is roughly
\begin{gather}
\frac{\omega_{p,0}^3K^2}{m n_0 \Omega_0}\,Q \sim
\left(\frac{eE_c}{mc\omega}\right)^2 \equiv a^2,
\end{gather}
where we assumed $Q \sim \omega^{-3}\int F_0\,d^3k \sim \mc{I}/\omega^3 \sim |E_c|^2/(8\upi\omega^4)$. Note that $a$ is the amplitude of the EM-driven momentum oscillations in units $mc$. Thus, in the nonrelativistic limit assumed here, one has $a \ll 1$, so the photon contribution is small. Nevertheless, photons can have an important effect on the LW stability. For example, when resonant photons are present, the integrand in \Eq{eq:Q} has a pole on the real axis, and the integral must be taken along the Landau contour \citep{book:stix}. Then, $\Omega$ is complex, which signifies dissipation (positive or negative) of LW on photons. This effect is known as photon Landau damping \citep{ref:bingham97, book:mendonca} and has been demonstrated experimentally, albeit in a solid-state medium rather than plasma \citep{ref:dylov08}. Also note that LW are considered here only as a simple example, and the effect of photons on the dispersion of other waves can be more substantial.

\section{Polarizability-based approach}
\label{sec:pol}

Although the calculations of the ponderomotive forces are relatively straightforward in the situation considered in this paper, the problem can be much harder when $\Phi_{\vec{k}}$ is velocity-dependent, i.e., when kinetic effects are essential \citep{my:itervar, my:trieste08, my:mneg}. Thus, it would be advantageous to develop a formulation of the modulational dynamics that avoids this step altogether. Below, we propose such formulation that utilizes the photon-polarizability concept \citep{tex:myqponder}. Within this approach, ponderomotive forces do not need to be considered, and a more general dispersion relation is obtained. 

\subsection{General theory}

The dispersion relation of electrostatic oscillations in plasma with dielectric tensor $\vec{\epsilon}$ is given by
\begin{gather}\label{eq:dr}
\vec{e}_{\vec{K}}\cdot \vec{\epsilon}(\Omega, \vec{K})\cdot \vec{e}_{\vec{K}}=0,
\end{gather}
where $\vec{e}_{\vec{K}}\doteq \vec{K}/K$ is the unit vector along the MW wave vector $\vec{K}$, and $\Omega$ is the MW frequency. Consider a plasma whose dielectric tensor is some $\vec{\epsilon}_0$ plus the contribution from the photon gas. The latter contribution is the photon susceptibility, which can be written as follows:
\begin{gather}
\vec{\chi}_{\rm ph}(\Omega, \vec{K}) = 4\upi \int \vec{\alpha}_{\rm ph}(\Omega, \vec{K}, \vec{k})\,f_{\rm ph}(\vec{k})\,d^3k.
\end{gather}
Here, $f_{\rm ph}$ is the unperturbed photon distribution, and $\vec{\alpha}_{\rm ph}$ is the polarizability of a single photon. As shown previously \citep{tex:myqponder},
\begin{gather}
\vec{\alpha}_{\rm ph}
= \frac{\hbar e^2K^2 \Xi}{4m^2\omega_0^3}\,\vec{e}_{\vec{K}}\vec{e}_{\vec{K}},
\end{gather}
where $\Xi$ is a dimensionless coefficient given by
\begin{gather}
\Xi \doteq \frac{\Omega^2 - c^2K^2}{(\Omega - \vec{K} \cdot \vec{v}_*)^2 - (\Omega^2-c^2K^2)^2/4\omega_0^2(k)},
\end{gather}
or, equivalently,
\begin{gather}
\Xi = - \sum_{\sigma =\pm 1} \frac{\sigma\omega_0(k)}{R_\sigma(\Omega, \vec{K}, \vec{k})},\\
R_\sigma(\Omega, \vec{K}, \vec{k}) \doteq \Omega -\vec{K} \cdot \vec{v}_*(\vec{k}) 
+ \sigma\,\frac{\Omega^2-c^2K^2}{2\omega_0(k)}.
\end{gather}
Thus, \Eq{eq:dr} can be expressed as follows:
\begin{gather}\label{eq:ep0c}
\epsilon_0(\Omega, \vec{K}) + \chi_{\rm ph}(\Omega, \vec{K}) = 0.
\end{gather}
Here, $\epsilon_0\doteq \vec{e}_{\vec{K}}\cdot \vec{\epsilon}_0(\Omega, \vec{K})\cdot \vec{e}_{\vec{K}}$, and $\chi_{\rm ph}(\Omega, \vec{K})\doteq \vec{e}_{\vec{K}}\cdot \vec{\chi}_{\rm ph}(\Omega, \vec{K})\cdot \vec{e}_{\vec{K}}$, or, explicitly,
\begin{gather}\notag
\chi_{\rm ph}(\Omega, \vec{K}) = - \frac{\omega_{p, 0}^2K^2}{4m n_0}  \sum_{\sigma =\pm 1} \int \frac{\sigma F_0(\vec{k})}{\omega_0^2(k) R_\sigma(\Omega, \vec{K}, \vec{k})}\,d^3k,
\end{gather}
where $F_0\doteq \hbar f_{\rm ph}$ is the photon action density, which is a classical quantity. Accordingly, \Eq{eq:ep0c} becomes
\begin{gather}\label{eq:gdr}
\epsilon_0(\Omega, \vec{K})-\frac{\omega_{p, 0}^2K^2}{4m n_0} 
 \sum_{\sigma =\pm 1} \int \frac{\sigma\, 
 F_0(\vec{k})}{\omega_0^2(k)R_\sigma(\Omega, \vec{K}, \vec{k})}\,d^3k = 0.
\end{gather}

Compared to \Eq{eq:Omg} that was obtained within a PDE-based approach, \Eq{eq:gdr} is more general. It allows for $k \sim K$ and does not assume any specific model of the background plasma; i.e., no assumptions regarding $\vec{\epsilon}_0$ are made. Moreover, the derivation applies as is to the case where the background plasma is weakly nonstationary and (or) weakly inhomogeneous. [For more details, see \citet{tex:myqponder}.] Also importantly, the same approach is readily extended to describe modulational dynamics of other waves too, as will be reported separately.

\subsection{Small-$K$ limit}

Finally, let us show how \Eq{eq:gdr} reduces to \Eq{eq:Omg} under the additional assumptions adopted in \Sec{sec:pde}. First, assume the GO approximation for photons, namely, $K \ll k$. This implies \citep{tex:myqponder, my:lens}
\begin{gather}
\Xi(\Omega, \vec{K}, \vec{k}) \approx \frac{\Omega^2-c^2K^2}{(\Omega - \vec{K} \cdot \vec{v}_*)^2}.
\end{gather}
Hence, the photon susceptibility becomes
\begin{gather}\notag
\chi_{\rm ph}(\Omega, \vec{K}) = \frac{\omega_{p, 0}^2K^2}{4m n_0} \int 
\frac{\Omega^2-c^2K^2}{(\Omega - \vec{K} \cdot \vec{v}_*)^2\omega_0^3(k)}\,F_0(\vec{k})\,d^3k.
\end{gather}
As can be checked by a direct calculation,
\begin{gather}
\vec{K} \cdot \frac{\pd}{\pd \vec{k}}\,
 \left[\frac{1}{(\Omega -\vec{K} \cdot \vec{v}_*)\,\omega_0^2}\right]
 = \frac{c^2K^2-\Omega^2}{(\Omega -\vec{K} \cdot \vec{v}_*)^2\,\omega_0^3} 
    + \frac{1}{\omega_0^3}.\label{eq:lemma}
\end{gather}
(This nonintuitive step can be avoided as explained in Appendix~\ref{app:var}.) Thus, one can also express $\chi_{\rm ph}$ equivalently as follows:
\begin{gather}
\chi_{\rm ph}(\Omega, \vec{K}) = 
\frac{\omega_{p, 0}^2K^2}{4m n_0} \,Q(\Omega, \vec{K}),
\end{gather}
where $Q$ is given by \Eq{eq:Q}. Second, assume that ions are motionless and electrons have a nonzero yet small enough thermal speed $v_T \ll \Omega/K$; then~\citep{book:stix},
\begin{gather}
\epsilon_0(\Omega, \vec{K}) = 1-\frac{\omega_{p, 0}^2}{\Omega^2}\Big(1 + \frac{3K^2v_T^2}{\omega_{p,0}^2}\Big),
\end{gather}
and $\Omega$ is close to the unperturbed linear frequency $\Omega_0$ [\Eq{eq:Om0}]. This can be simplified to $\epsilon_0 \approx 2[\Omega -\Omega_0(\vec{K})]/\omega_{p, 0}$. Hence, \Eq{eq:ep0c} immediately leads to \Eq{eq:Omg}.

\section{Conclusions}

In summary, we calculated the influence of the photon polarizability on the dispersion of linear Langmuir waves in collisionless nonmagnetized electron plasma. Two alternative formulations are given here. In the first formulation, we calculate the Langmuir-wave dispersion by solving the equations of motion for the electron fluid, where the photon contribution enters as a ponderomotive force. A related calculation was reported previously \citep{ref:bingham97, book:mendonca}, but it has omissions, which are corrected in the present work. In the second formulation, we explicitly invoke the photon-polarizability concept \citep{tex:myqponder}. Then, the theory becomes linear, ponderomotive forces do not need to be considered, and,  consequently, more general results are obtained.

Although we focus on LW, the calculation presented here is only a characteristic example. The concept of the photon (plasmon, phonon, etc.) polarizability is of broader generality and can help calculate the modification of the dispersion relation also of other waves, including waves in media other than plasma. We leave the consideration of specific examples to future publications.

The work was supported by the NNSA SSAA Program through DOE Research Grant No. DE-{NA0002948}, by the U.S. DOE through Contract No. DE-AC02-09CH11466, and by the U.S. DOD NDSEG Fellowship through Contract No. 32-CFR-168a. 

\appendix
\section{Variational approach}
\label{app:var}

Here, we present an even more straightforward derivation of the GO dispersion relation \eq{eq:Omg} that does not involve the nonintuitive step of using \Eq{eq:lemma}. Instead of invoking the photon-polarizability concept, we start directly with the LW Lagrangian density~\citep{tex:myqponder}:
\begin{gather}\label{eq:Lcc}
\mcc{L} = \epsilon_0(\Omega, \vec{K})\,\frac{|\tilde{\vec{E}}_c|^2}{16\upi} - \int \Phi_{\rm ph} f_{\rm ph}\,d^3k.
\end{gather}
Here, the first term is the Lagrangian density of LW absent photons \citep{my:itervar}, the second one is the photon contribution, and $\Phi_{\rm ph}$ is the LW-produced ponderomotive potential \textit{of a photon}. Let us use the GO approximation of $\Phi_{\rm ph}$ that was derived in \citet{tex:myqponder, my:lens}:
\begin{gather}
\hbar^{-1}\Phi_{\rm ph} = \langle \omega - \omega_0 \rangle + \frac{\vec{K}}{4}\cdot
\frac{\pd}{\pd \vec{k}}
\left(
\frac{|\tilde{\omega}_c|^2}{\Omega - \vec{K} \cdot \vec{v}_*}
\right).
\end{gather}
(The angular brackets denote averaging over time.) As seen from \Eq{eq:om},
\begin{gather}
\langle \omega - \omega_0 \rangle = - \frac{\omega_{p,0}^4}{16\omega_0^3}\,\frac{|\tilde{n}_c|^2}{n_0^2}.
\end{gather}
Also, from Gauss's law, $\tilde{n}_c = i \vec{K} \cdot \tilde{\vec{E}}_c/(4\upi e)$, so
\begin{gather}
\frac{|\tilde{n}_c|^2}{n_0^2} = \frac{K^2 |\tilde{\vec{E}}_c|^2}{16\upi^2n_0^2e^2} 
= \frac{|\tilde{\vec{E}}_c|^2}{16\upi} \,\frac{4K^2}{mn_0 \omega_{p,0}^2},
\end{gather}
where we used that the electrostatic field $\tilde{\vec{E}}_c$ is parallel to $\vec{K}$. This gives
\begin{gather}\notag
\hbar^{-1}\Phi_{\rm ph} = 
\frac{\omega_{p,0}^2 K^2}{4 mn_0} \left\{-\frac{1}{\omega_0^3} +
\vec{K} \cdot
\frac{\pd}{\pd \vec{k}}
\left[
\frac{1}{(\Omega - \vec{K} \cdot \vec{v}_*)\omega_0^2}
\right]\right\},
\end{gather}
where we used \Eq{eq:omc}. By substituting this result into \Eq{eq:Lcc} and integrating by parts, one obtains
\begin{gather}\label{eq:Lf}
\mcc{L} = \epsilon(\Omega, \vec{K})\,\frac{|\tilde{\vec{E}}_c|^2}{16\upi},
\end{gather}
where $\epsilon$ is given by
\begin{gather}
\epsilon(\Omega, \vec{K}) \doteq \epsilon_0(\Omega, \vec{K}) 
+ \frac{\omega_{p,0}^2 K^2}{4 mn_0}\, Q(\Omega, \vec{K}),
\end{gather}
and $Q$ is given by \Eq{eq:Q}. The corresponding action integral can be considered as a functional of $\mc{A} \doteq |\tilde{\vec{E}}_c|^2$ and the LW phase $\Theta$; namely,
\begin{gather}
S = \int 
\mcc{L}(\mc{A}, \underbrace{-\pd_t\Theta}_\Omega, \underbrace{\nabla\Theta\vphantom{-\pd_t\Theta}}_{\vec{K}})\,dt\,d^3x.
\end{gather}
The dispersion relation is obtained from $\delta S/\delta \mc{A} = 0$. This gives $\epsilon(\Omega, \vec{K}) = 0$, so one is again led to \Eq{eq:Omg}. Also notably, if $\epsilon$ depends on $(t, \vec{x})$, the amplitude equation (action conservation theorem) is obtained from $\delta S/\delta \Theta = 0$; namely,
\begin{gather}\label{eq:act}
\pd_t \mc{I} + \nabla \cdot (\vec{V}_g \mc{I}) = 0.
\end{gather}
Here, $\mc{I} = (\pd_\Omega\epsilon)|\tilde{\vec{E}}_c|^2/(16\upi)$ is the LW wave action density, $\vec{V}_g \doteq -(\pd_{\vec{K}}\epsilon)/(\pd_\Omega\epsilon)$ is the LW group velocity, $\Omega$ is found from the dispersion relation, and the wave vector is treated as a field; i.e., $\vec{K} = \vec{K}(t, \vec{x})$ \citep{my:wkin, book:tracy}.

Although this variational formalism does not capture dissipation [which is reflected in the conservative form of \Eq{eq:act}], it can be extended to dissipative processes as described in \citet{tex:mynonloc}. In particular, the dispersion relation derived from the above variational formalism is valid for complex frequencies too provided that the integral in the expression for $Q$ [\Eq{eq:Q}] is taken using the Landau rule.

\end{document}